\documentclass[10pt, conference]{IEEEtran}
\usepackage{psfrag}
\usepackage[dvips]{graphicx}
\DeclareGraphicsExtensions{.eps}
\ifCLASSINFOpdf
\else
\fi
\usepackage[cmex10]{amsmath}

\begin{document}
%
\title{Outage Probability of Power-based \\ Non-Orthogonal Multiple Access (NOMA) \\
on the Uplink of a 5G Cell}

\author{\IEEEauthorblockN{
 Maria Luisa Merani
}
\IEEEauthorblockA {
Dipartimento di Ingegneria ``Enzo Ferrari''\\
University of Modena and Reggio Emilia,%
Modena, Italy\\
{e-mail:marialuisa.merani@unimore.it}
}
}


%


\maketitle

\begin{abstract}

This letter puts forth an analytical approach to evaluate the outage probability  of power-based NOMA on the uplink of a 5G cell, the outage being defined as the event where the receiver fails to successfully decode all the simultaneously received signals.
In the examined scenario, Successive Interference Cancellation (SIC) is considered and an arbitrary number of superimposed signals is present. 
For the Rayleigh fading case, the outage probability is provided in closed-form, clearly outlining its dependency on the signal-to-noise ratio of the users that are simultaneously transmitting, as well as on their distance from the receiver.
\end{abstract}

\noindent
{\bf Index Terms - Uplink NOMA, successive interference cancellation, outage probability, 5G} 

\noindent
{\bf IEEE CL EDICS: CL1.2.0, CL1.2.1}

%
\IEEEpeerreviewmaketitle

%

\section{Introduction}
Non-orthogonal Multiple Access has in recent years stirred a great deal of interest, because of its promise of enhancing the capacity of 5G cellular systems. 
NOMA allows different simultaneous users to share the available system resources (frequency, time) through a variety of different techniques, well illustrated in \cite{SPECIAL_TOPIC} and \cite{CHIN_LI}: as a matter of fact, NOMA can operate in the power-domain, can adopt spreading sequences, can rely on coding matrices and/or interleaving.
This letter focuses on power-based NOMA on the uplink of a 5G cell, when a SIC receiver is employed. 
In \cite{COMM_LETT_CINESI}, emphasis was on uplink power-based NOMA too, the goal being to evaluate the achievable sum data rate and the corresponding outage, which was provided in closed-form for the case of two users.
The latter condition is commonly encountered in literature, as it guarantees a tractable analysis. Unlike \cite{COMM_LETT_CINESI} and previous works, this letter sets
no limit  to the number of superimposed signals. 
Furthermore, mediating from \cite{EKRAM}, an alternative approach is taken with respect to the outage probability, which is defined as the probability that the recovery of all signals fails, given the constraint on the received powers required by SIC is not observed. Lastly, for the case of Rayleigh fading, the developed theoretical analysis provides the outage probability in closed-form, immediately revealing the influence of several factors, among which frequency, distance and different signal-to-noise ratio assignments, on system performance. The numerical results explore the setting where either two o three superimposed signals are present, and show that several configurations exist, where the outage is confined to low values and the benefit of NOMA can indeed be effectively exploited.  

The remainder of the letter is organized as follows: Section II defines the scenario of investigation and develops the theoretical analysis; the case of Rayleigh fading is illustrated in Section III; Section IV provides some reference results, while the conclusions are drawn in Section V.

\section{Scenario and Performance Analysis}
Within the current work, uplink communications in a 5G cell are examined; power-based NOMA is employed and the  reference scenario features $n$ $UE$'s (User Equipments) that transmit to the enodeB, the enhanced node B, on the
same radio spectrum.
Let $h_i$ denote the envelope of the channel between the $i$-th $UE$, $UE_i$, and enodeB, $i=1,2,\ldots,n$, and
as in \cite{EKRAM}, let $\Gamma_i=\frac{h_i^2}{N_{0}B}$ indicate the $i$-th normalized channel gain, $N_{0}$ being the noise power spectral density and $B$ the transmission bandwidth; further, let
$
p_{\Gamma_i}(\gamma_i) 
$
be 
the  probability density function (pdf) of the generic $\Gamma_i$.
The assumption is that $\Gamma_i$ and $\Gamma_j$ are independent random variables with different mean values, $\forall i$ and $j$. 
Unlike LTE, 5G uplink power-based NOMA associates different 
transmit power levels  to different $UE$'s, in order to guarantee different received powers at the enodeB and to facilitate the task of the interference cancellation receiver.
Namely, the more favorable the channel gain that the $UE$ experiences, the higher the power level that the $UE$ works with.
It follows that the instantaneous values of the $\Gamma_i$'s have to be ordered, so as to obtain
$\Gamma_{(1)}$, $\Gamma_{(2)}$, $\ldots$, $\Gamma_{(n)}$, with
\begin{equation}
\Gamma_{(1)} > \Gamma_{(2)} > \ldots > \Gamma_{(n)}
\, ,
\label{GAMMAS}
\end{equation}
where these new random variables are no longer independent;
then, transmit powers are assigned to $UE$'s respecting the constraint 
\begin{equation}
P_{(1)} > P_{(2)} > \ldots > P_{(n)}
\, ,
\label{PIS}
\end{equation}
$P_{(i)}$ representing the transmit power of the $UE$ with the $i$-th largest channel gain $\Gamma_{(i)}$,
denoted by $UE_{(i)}$.
The enodeB receives the superimposed messages from the $UE$'s  and through SIC it decodes the strongest signal first, then the second strongest, until the last.  

We are interested in the occurrence of unsuccessful decoding and begin by observing that the strongest signal received by the enodeB from $UE_{(1)}$ cannot be detected if
\begin{equation}
P_{(1)}\Gamma_{(1)}-P_{(2)}\Gamma_{(2)}-P_{(3)}\Gamma_{(3)}-\ldots-P_{(n)}\Gamma_{(n)}<P_{thres}
\, ,
\label{out_condition}
\end{equation}
where $P_{thres}$ indicates the minimum power difference which allows to successfully extract the first useful signal.
If last inequality holds, then the second strongest received signal from $UE_{(2)}$ cannot be decoded either; 
as a matter of fact, its recovery with no cancellation of the signal received from $UE_{(1)}$ would require 
\begin{equation}
-P_{(1)}\Gamma_{(1)}+P_{(2)}\Gamma_{(2)}-P_{(3)}\Gamma_{(3)}-\ldots-P_{(n)}\Gamma_{(n)}>P_{thres}
\, ,
\end{equation}
which owing to the $P_{(1)}\Gamma_{(1)}>P_{(2)}\Gamma_{(2)}$ condition can never be satisfied.
In turn, signal from $UE_{(3)}$ cannot be decoded either, without the prior successful decoding of signal from $UE_{(2)}$.
The final consequence of this iterated reasoning is that inequality
(\ref{out_condition}) represents the outage condition, that is, the condition when none among the superposed signals can be correctly recovered.
Hence, the outage probability of power-based NOMA in the presence of $n$ simultaneous transmissions is defined as 
\begin{equation}
P_{out_n}=Pr\{P_{(1)}\Gamma_{(1)}-\sum_{i=2}^{n}P_{(i)}\Gamma_{(i)}<P_{thres}\}
\, .
\label{outage_probability}
\end{equation}

Evaluating (\ref{outage_probability}) requires the consideration of $n$ dependent random variables, the generic of which is 
\begin{equation}
X_{(i)}=P_{(i)}\Gamma_{(i)}
\, ;
\label{ordered_Xi}
\end{equation} 
recalling both (\ref{GAMMAS})  and (\ref{PIS}), it is immediate to conclude that 
condition $X_{(1)}\ge X_{(2)}\ge \ldots \ge X_{(n)}$ holds.

It is now instructive to begin by considering $n=2$ and to indicate by $f_{joint_2}(x_{(1)},x_{(2)})$ the joint pdf of $X_{(1)}$ and $X_{(2)}$, so that 
the outage probability in (\ref{outage_probability}) becomes
\begin{eqnarray}
& P_{out_2}=Pr\{P_{(1)}\Gamma_{(1)}-P_{(2)}\Gamma_{(2)}<P_{thres}\}= \nonumber \\
&=Pr\{X_{(1)}-X_{(2)}<P_{thres}\}= \nonumber \\
& =1-Pr\{X_{(2)}<X_{(1)}-P_{thres} \}=
\nonumber \\
&=1-\int_{x_{(1)}=P_{thres}}^{\infty}
\int_{x_{(2)}=0}^{x_{(1)}-P_{thres}}
f_{joint_2}(x_{(1)},x_{(2)})
\nonumber \\
& dx_{(2)}dx_{(1)}
\, .
\label{pout_2}
\end{eqnarray}
For a generic $n$, last expression generalizes to
\begin{eqnarray}
& P_{out_n}
=1-\int_{x_{(1)}=P_{thres}}^{\infty}
\int_{x_{(2)}=0}^{x_{(1)}-P_{thres}}
\int_{x_{(3)}=0}^{x_{(1)}-x_{(2)}-P_{thres}}
\ldots
\nonumber \\
& \int_{x_{(n)}=0}^{x_{(1)}-x_{(2)}-x_{(3)}-\ldots-x_{(n-1)}-P_{thres}}
f_{joint_n}(x_{(1)},x_{(2)}, \ldots,x_{(n)}) \cdot
\nonumber \\
& \cdot dx_{(n)} dx_{(n-1)}\ldots dx_{(1)}
\, , 
\label{pout_n}
\end{eqnarray}
where $f_{joint_n}(x_{(1)},x_{(2)}, \ldots,x_{(n)})$ indicates the joint pdf of 
the ordered set $X_{(1)}$, $X_{(2)}$, $\ldots$, $X_{(n)}$. 
Now the problem at hand is to determine this pdf. 
In this respect,
let $F$ be the $n\times n$ matrix defined as
\begin{equation}
F= 
\left [
\begin{array}{llll}
f_1(x_{(1)}) & f_2(x_{(1)}) & \ldots & f_n(x_{(1)}) \\
f_1(x_{(2)}) & f_2(x_{(2)}) & \ldots & f_n(x_{(2)}) \\
\vdots   &  \vdots  & \ddots & \vdots   \\
f_1(x_{(n)}) & f_2(x_{(n)}) & \ldots & f_n(x_{(n)}) \\
\end{array}
\right ]
\, ,
\end{equation}
$f_i(\cdot)$ being the pdf of the -- unordered -- random variable $X_{i}$, defined as
\begin{equation}
X_i=P_{(i)}\cdot \Gamma_i \, , \qquad i=1,2,\dots,n
\, ,
\label{Xi}
\end{equation}
whose pdf is immediately determined, once $p_{\Gamma_i}(\gamma_i)$ is known, as $P_{(i)}$ is a constant. 
Next, recall that the permanent of a square matrix $A$, written as $\stackrel{+}{|}A\stackrel{+}{|}$, is defined like the determinant, except that all signs are positive. 
Then, the joint pdf $f_{joint_n}(x_{(1)},x_{(2)},\ldots,x_{(n)})$ can be expressed as
\begin{equation}
f_{joint_n}(x_{(1)},x_{(2)},\ldots,x_{(n)})=\stackrel{+}{|}F\stackrel{+}{|}
\, .
\label{joint_pdf}
\end{equation}
Last result is substantiated by the reasoning in \cite{VAUGH}, where the arguments of \cite{KENDALL} are extended to prove 
the formulation in (\ref{joint_pdf}) with the use of permanents.

At first sight, $f_{joint_n}(x_{(1)},x_{(2)},\ldots,x_{(n)})$ gives the impression that evaluating the integral in (\ref{pout_n}) might be quite cumbersome. However, the joint pdf obeys a highly peculiar structure and an alike -- and more convenient -- writing of it is provided in the following terms: 
let $S_N$ indicate the group of all $n!$ permutations of the set $N=\{1,2,\ldots,n\}$ and by $S_i=\{i_1,i_2,\ldots,i_n\}$ the generic of such permutations. It follows that  $f_{joint_n}(x_{(1)},x_{(2)},\ldots,x_{(n)})$ is equivalently written as
\begin{eqnarray}
f_{joint_n}(x_{(1)},x_{(2)},\ldots,x_{(n)})=
\nonumber \\
=\sum_{S_i\in S_N}f_{1}(x_{(i_1)})f_{2}(x_{(i_2)})\cdot\ldots\cdot f_{n}(x_{(i_n)})
\, .
\label{permutations}
\end{eqnarray}
Last expression highlights that the joint pdf exhibits the presence of $n!$ terms, wherein the permutations of the arguments of the $f_1(\cdot)$, $f_2(\cdot)$, $\ldots$, $f_n(\cdot)$ pdf's appear. 
Hence, if we indicate by $I_{S_i}$ the result of the integral
\begin{eqnarray}
& I_{S_i}=
\int_{x_{(i_1)}=0}^{\infty}
\int_{x_{(i_2)}=0}^{x_{(i_1)}-P_{thres}}
\int_{x_{(i_3)}=0}^{x_{(i_1)}-x_{(i_2)}-P_{thres}}
\ldots
\nonumber \\
& \int_{x_{(i_n)}=0}^{x_{(i_1)}-x_{(i_2)}-x_{(i_3)}-\ldots-x_{i_{(n-1)}}-P_{thres}}
\nonumber \\
& p_{i_1 i_2 \ldots i_n}(x_{(i_1)},x_{(i_2)},\ldots, x_{(i_n)}) \cdot 
\nonumber \\
& \cdot dx_{(i_n)} dx_{i_{(n-1)}}\ldots dx_{(i_1)}
\, ,
\label{IntegralSi}
\end{eqnarray}
where 
\begin{eqnarray}
p_{i_1 i_2 \ldots i_n}(x_{(i_1)},x_{(i_2)},\ldots, x_{(i_n)})=
\nonumber \\
f_{1}(x_{(i_1)})f_{2}(x_{(i_2)})\cdot\ldots\cdot f_{n}(x_{(i_n)})
\, ,
\label{p_i1i2in}
\end{eqnarray}
then we observe that
\begin{equation}
P_{outn}=1-\sum\limits_{S_i\in S_N} I_{S_i}
\, .
\end{equation}
Luckily, when the random variables $X_{1}$, $X_{2}$, $\ldots$, 
$X_{n}$ obey the same statistical description, 
although with different mean values, for a permutation $S_j$ different than $S_i$, the $I_{S_j}$ result is readily obtained from $I_{S_i}$ 
through the analogous permutation of the $f_i(\cdot)$'s in (\ref{p_i1i2in}), $\forall j$. 
That is to say, given the 
$n$-th fold integral in (\ref{IntegralSi}) has been solved once, e.g., $I_{S_1}$ has been determined, $S_1=\{1,2,\ldots,n\}$, then all the 
remaining $I_{S_i}$'s are known.
As an illustrative example, the case of Rayleigh fading is examined next.

\section{Rayleigh fading case}
When the envelope of the received signal is subject to Rayleigh fading, $\Gamma_i$ and in turn $X_{i}$ are exponentially distributed
with means $\overline \Gamma_i$ and $\overline X_{i}$, respectively.
Beginning with the case $n=2$, $P_{out_2}$ is expressed by
\begin{equation}
P_{out_2}=1-(I_{S_1}+I_{S_2})
\, ,
\end{equation}
where $S_1=\{1,2\}$ and 
\begin{equation}
I_{S_1}=\int\limits_{x_{(1)}=P_{thres}}^{\infty}
\int\limits_{x_{(2)}=0}^{x_{(1)}-P_{thres}}
p_{12}(x_{(1)},x_{(2)}) dx_{(2)}dx_{(1)}
\label{IS1}
\end{equation} 
with  
\begin{eqnarray}
p_{12}(x_{(1)},x_{(2)})= f_1(x_{(1)})f_2(x_{(2)})= \nonumber \\
= \frac{1}{\overline X_1} exp\left ({-\frac{x_{(1)}}{\overline X_{1}}} \right )\cdot
\frac{1}{\overline X_{2}} exp\left ({-\frac{x_{(2)}}{\overline X_{2}}} \right )
\, ;
\label{p12}
\end{eqnarray} 
analogously, $S_2=\{2,1\}$ and 
\begin{equation}
I_{S_2}=\int\limits_{x_{(1)}=P_{thres}}^{\infty}
\int\limits_{x_{(2)}=0}^{x_1-P_{thres}}
p_{21}(x_{(1)},x_{(2)}) dx_{(2)}dx_{(1)}
\end{equation} 
with 
\begin{eqnarray}
p_{21}(x_{(1)},x_{(2)})= f_1(x_{(2)})f_2(x_{(1)})= \nonumber \\
= \frac{1}{\overline X_{1}} exp\left ({-\frac{x_2}{\overline X_{1}}} \right )\cdot
\frac{1}{\overline X_{2}} exp\left ({-\frac{x_1}{\overline X_{2}}} \right )
\, .
\label{p21}
\end{eqnarray} 

Solving the integral in (\ref{IS1}) gives
\begin{equation}
I_{S_1}=\frac{\overline X_1}{\overline X_1+\overline X_2}
exp\left (-
\frac{P_{thres}}{\overline X_1}
\right)
\, ,
\end{equation}
wherefore $I_{S_2}$ immediately follows as
\begin{equation}
I_{S_2}=\frac{\overline X_2}{\overline X_2+\overline X_1}
exp\left (-
\frac{P_{thres}}{\overline X_2}
\right)
\, ,
\end{equation}
and finally
\begin{eqnarray}
& P_{out2}=1+
\\ \nonumber
& -\left ( \frac{\overline X_1}{\overline X_1+\overline X_2}
exp\left (
\frac{-P_{thres}}{\overline X_1}
\right)+
\frac{\overline X_2}{\overline X_1+\overline X_2}
exp\left (
\frac{-P_{thres}}{\overline X_2}
\right)
\right )
\, .
\end{eqnarray}

Similarly, when $n=3$, there will be $3!$ distinct integral contributions of the type in (\ref{IntegralSi}) in the outage probability expression, that are determined once $p_{123}(x_{(1)},x_{(2)},x_{(3)})$ is introduced,
\begin{eqnarray}
&p_{123}(x_{(1)},x_{(2)},x_{(3)})=
\nonumber \\
&\frac{1}{\overline{X_{1}}} exp\left ({-\frac{x_{(1)}}{\overline X_{1}}} \right )\cdot
\frac{1}{\overline X_{2}} exp\left ({-\frac{x_{(2)}}{\overline X_{2}}} \right )\cdot
\nonumber \\
& \cdot \frac{1}{\overline X_{3}} exp\left ({-\frac{x_{(3)}}{\overline X_{3}}} \right )
\, .
\end{eqnarray}
Now, $S_1=\{1,2,3\}$ and
\begin{eqnarray}
& I_{S_1}=\int_{x_{(1)}=P_{thres}}^{\infty}
\int_{x_{(2)}=0}^{x_{(1)}-P_{thres}}
\int_{x_{(3)}=0}^{x_{(1)}-x_{(2)-P_{thres}}}
\nonumber \\
&
p_{123} (x_{(1)},x_{(2)},x_{(3)}) \cdot 
\nonumber \\ 
& \cdot dx_{(3)} dx_{(2)} dx_{(1)}  = \frac
{\overline{X_1}^2 exp(-\frac{P_{thres}}{\overline X_1})}
{(\overline{X_1}+\overline{X_2})(\overline{X_1}+\overline{X_3})}
\, ,
\end{eqnarray}
so that, after a few passages, $P_{out_3}$ is determined as
\begin{eqnarray}
P_{out_3}=1-\sum_{S_i\in S_N}I_{S_i}=
\nonumber \\
1- \frac{3!}{3}
\left (
\frac{\overline X_1^2 exp\left (-\frac{P_{thres}}{\overline X_1}\right )}{(\overline X_1+\overline X_2)(\overline X_1+\overline X_3)}
+
\right.
\nonumber \\
+
\frac{\overline X_2^2 exp\left (-\frac{P_{thres}}{\overline X_2}\right)}{(\overline X_2+\overline X_1)(\overline X_2+\overline X_3)}
+ 
\nonumber \\
+
\left.
\frac{\overline{X_3}^2 exp\left ( -\frac{P_{thres}}{\overline X_3}\right )}{(\overline X_3+\overline X_1)(\overline X_3+\overline X_2)}
\right )
\, .
\end{eqnarray}
Iterating the procedure, by induction it is proved that $P_{out_n}$, the outage probability in the presence of $n$ superposed signals, is given in closed form by:
\begin{eqnarray}
P_{out_n}
=
1-
\frac{n!}{n}\cdot
\sum_{k=1}^{n}
\frac
{\overline X_k^{n-1}e^{-\frac{P_{thresh}}{\overline X_k}}}
{\prod\limits_{\stackrel{i=1}{i \neq k}}^{n}(\overline X_k+\overline X_i)}=
\nonumber \\
=1-(n-1)! \sum_{k=1}^{n}
\frac
{e^{-\frac{P_{thres}}{\overline X_k}}}
{\prod\limits_{\stackrel{i=1}{i \neq k}}^{n}(1+\frac{\overline X_i}{\overline X_k})}
\, .
\label{Poutn}
\end{eqnarray}

Last expression allows to determine the outage probability in the presence of an arbitrary number of superimposed signals in a very effective and quick manner. 
To this regard, it is observed from (\ref{Xi}) that 
\begin{equation}
\overline X_i=P_{(i)}\overline \Gamma_i=
\frac{P_{(i)}}{N_{0}B}\cdot  \overline{h_i^2} =SNR_{(i)} \cdot \overline{h_i^2}
\end{equation}
where $SNR_{(i)}$ is the signal-to-noise ratio of $UE_{(i)}$.
Hence, when the $SNR_{(i)}$'s and the $\overline{h_i^2}$'s are provided, the outage probability of uplink power-based NOMA is immediately known.
In next section, some numerical examples will be offered, assuming that the path loss is such that
\begin{equation}
\overline {h_i^2}=
k_p \cdot D_{i}^{-\alpha}
\end{equation}
where  $D_{i}$ the distance between  the $i$-th $UE$ and enodeB, $\alpha$ the decay factor and $k_p$ is
$k_p=(\frac{c}{4\pi f_c})^2$, $c$ being the speed of light, $f_c$ the operating frequency and isotropic antennas being considered.
In this circumstance, (\ref{Poutn}) specializes to
\begin{equation}
P_{out_n}=1
- (n-1)!
\sum\limits_{k=1}^{n}
\frac
{exp \left ( -\frac{P_{thres}}{SNR_{(k)} k_p D_{k}^{-\alpha}}\right )}
{
\prod\limits_{\stackrel{i=1}{i\neq k}}^{n}
\left ( 
1+\frac{SNR_{(i)}}{SNR_{(k)}}\cdot \left( \frac{D_{i}}{D_{k}}\right )^{-\alpha}
\right )
}
\, .
\label{eq:pout-d}
\end{equation}
Given $P_{thres}$ is fixed, as well as the operating frequency $f_c$, the set of distances $D_{1}$, $D_{2}$, $\ldots$, $D_{n}$ and the SNR values $SNR_{(1)}$, $SNR_{(2)}$, $\ldots$, $SNR_{(n)}$, the probability of not being able to take advantage of successive interference cancellation is determined right away.
Next Section relies on (\ref{eq:pout-d}) to offer some meaningful insights on the performance of power-based NOMA employed in conjunction with SIC.
%
%
%
%
%

\section{Numerical Results}
\begin{figure}
\centering
\includegraphics[width=9truecm]{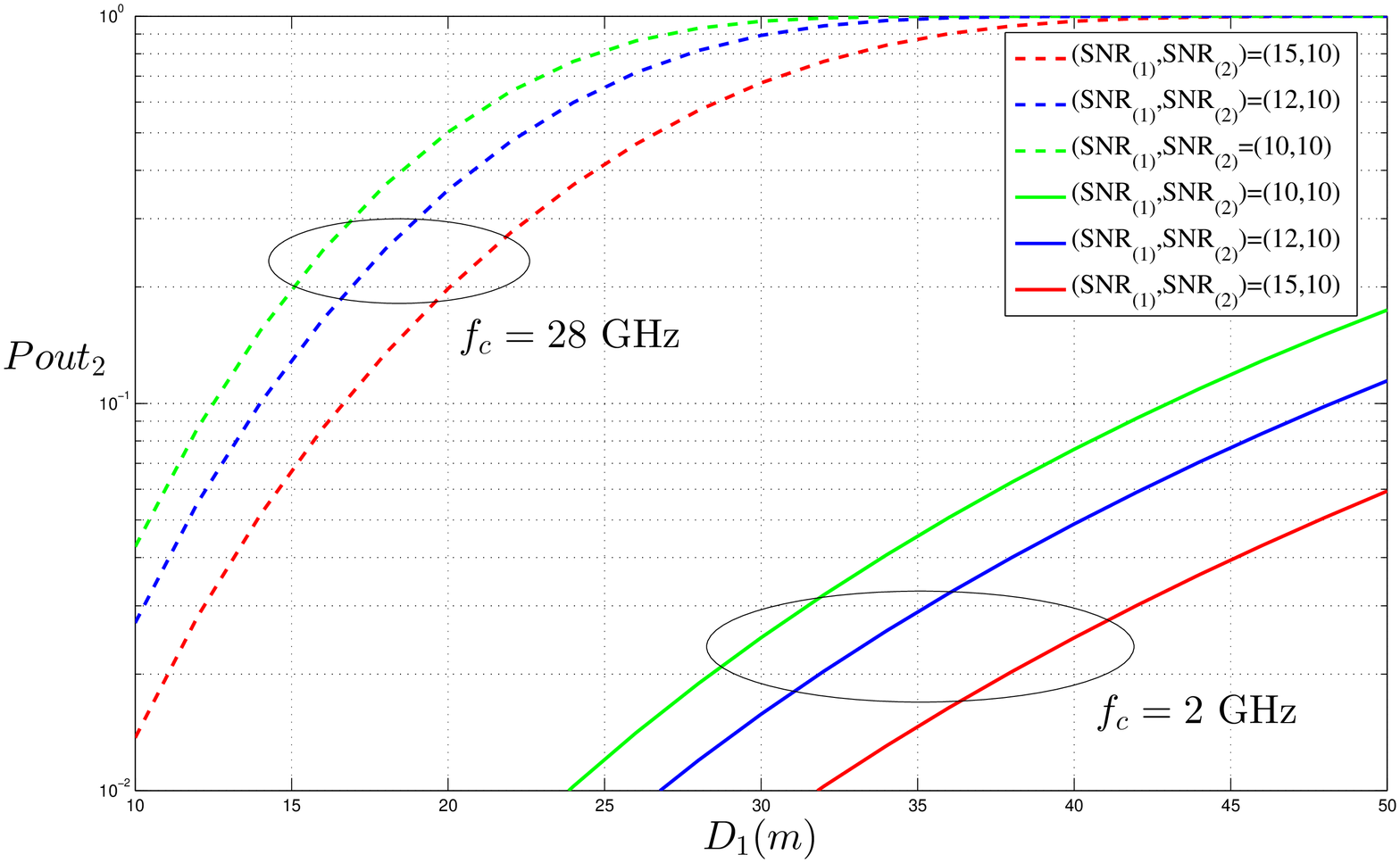}
\caption{$P{out_2}$ as a function of $D_{1}$, $f_c=2$ and $f_c=28$ GHz}
\label{fig:pout2_d}
\end{figure}

\begin{figure}[htb]
\centering
\includegraphics[width=9truecm]{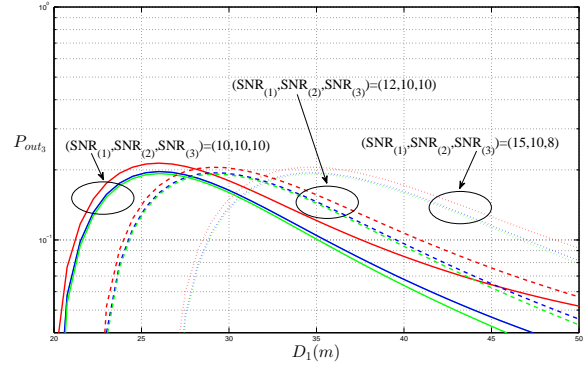}
\caption{$P{out_3}$ as a function of $D_{1}$, $f_c=28$ GHz}
\label{fig:pout3_d}
\end{figure}

Fig.\ref{fig:pout2_d} reports the outage probability in the presence of two superimposed signals as a function of $D_{1}$, the distance of $UE_{1}$ from the enodeB given in meters, when the second $UE$, $UE_{2}$, is at the cell edge, the cell radius is $R=100$ m, $SNR_{(1)}$ takes on different values, namely, $=11,8$ and $6$ dB, whereas $SNR_{(2)}=6$ dB. 
That is to say, a difference of $5$, $3$ and $0$ dB between the transmitted powers of the two $UE$'s is considered; this is in line with the choices performed in \cite{COMM_LETT_CINESI}, where the transmitted powers of two simultaneous users differ
for either $3$ or $5$ dB.
The propagation factor is $\alpha=4$ and two distinct values of the carrier frequency are considered: $f_c=2$ GHz (solid lines) and $28$ GHz (dashed lines). $P_{thres}$, the minimum difference in received powers is equal to $-75$ dBm \cite{RAPPA}.
The frequency effect on the outage probability is evident, highlighting that power-based NOMA is by far more attractive at lower frequencies.  
Nevertheless, interesting outage values can be attained when the distance of $UE_{1}$ from the enodeB is small and the gap between $SNR_{(1)}$ and $SNR_{(2)}$ increases.
Fig.\ref{fig:pout3_d} extends the reasoning to the case of three superimposed signals and shows the behavior of the outage probability as a function of $D_{1}$ given in meters for three distinct choices of the $(SNR_{(1)},SNR_{(2)},SNR_{(3)})$ triplet, namely: $(10,10,10)$ (solid lines),$(12,10,8)$ (dashed lines) and $(15,10,8)$ (dotted lines), when the carrier frequency is $f_c=28$ GHz. Moreover, different locations of $UE_{2}$ and $UE_{3}$ are examined: $D_{2}=0.2R$ paired with $D_3=0.5R$ (red lines), 
$D_{2}=0.2R$ with $D_{3}=0.7R$ (blue lines), and $D_{2}=0.2R$ with $D_{3}=0.9R$ (green lines), where it is recalled that the $R$ symbol indicates the cell radius.
Interestingly, all curves exhibit a similar shape; however, more pronounced differences in the SNR's have the effect of widening
the range of $D_1$ values for which the outage probability falls below a predefined limit (e.g., $10^{-1}$, $10^{-2}$).
Overall, note that the outage probability values are definitely worth of interest.
Moreover, the advantage of markedly separating the $UE$'s in terms of  $SNR$ values,
assigning the $UE$'s with the most favorable channel a higher $SNR$ value is manifest and numerically quantified. 
 
%
\section{Conclusions}
This paper has identified a novel, analytical method to  determine the probability of not being able to  take advantage of power-based NOMA on the uplink of a 5G cell, when successive interference cancellation  is employed and an arbitrary number of superimposed signals are considered. 
As a representative example, Rayleigh fading has been examined and the corresponding outage probability determined in closed-form.
The outage probability dependency on the carrier frequency, the signal-to-noise ratio of the $UE$'s that are simultaneously transmitting, as well as on their distance from enodeB
has been clearly pointed out, revealing that there exist several operating regions where power-based NOMA combined with SIC exhibits notably low outage probability values, even in the presence of several simultaneous users.

\end{document}